# From ATM to MPLS and QCI: the evolution of Differentiated QoS (D-QoS) standards and implications for 5G Network Slicing

Emeka Obiodu*, Nishanth Sastry†
* Department of Engineering, King's College London, UK. Email: chukwuemeka.obiodu@kcl.ac.uk

† Department of Computer Science, University of Surrey, UK. Email: n.sastry@surrey.ac.uk

*Abstract*—The networking community continues to create new technologies, and update existing ones to improve the quality, reliability and 'tailorability' of data networks. However, whenever Internet Service Providers (ISPs) attempt to productize 'tailorability' and sell it explicitly to end customers as a premium service over the 'best effort' connectivity, they either fail to overcome net neutrality concerns, or struggle to gain market traction. For this article, we focus only on those networking protocols, technologies or standards whose goal is to offer tailored connectivity to paying customers on a public network and refer to them as Differentiated QoS (D-QoS) standards. This article makes two contributions to the understanding of the goals of D-QoS standards. First, it explores their techno-economic market trajectory to understand the factors that determine success. In doing this, we acknowledge that, while there is wide variation and dissimilarity in their underlying technical properties, the expectation and goal for all D-QoS standards is that they will be used to provide a guaranteed connection that customers could be prepared to pay for. As such, we consider transport layer technologies (e.g. ATM, Frame Relay, MPLS), signaling technologies (e.g. RSVP), data packet markers (e.g. IP ToS, DiffServ, WME, QCI), and end-to-end separation solutions (e.g. Leased lines, Network Slicing) as a single cohort and analyse them together. Second, by exploring the parallels with 5G Network Slicing, we argue that despite its inherent technical differences with other D-QoS standards, the commercial performance of Network Slicing may end up resembling that of previous D-QoS standards. Consequently, we seek to learn lessons from previous D-QoS attempts and suggest that enterprise-focused 5G slices, running within a single service provider's domain and with binding service level agreements, will have the highest chance of success in the short/medium term.

## I. INTRODUCTION

The remarkable success of the internet and telecommunication networks relies on the numerous networking standards developed by the technical community in the past 50 years. These standards have steadily improved the quality and reliability of connectivity, and occasionally offered the possibility for tailoring the quality of the 'best effort' connectivity for different users or use cases. Among these standards, there is a subset which were designed, whether explicitly or implicitly, to provide a form of quality of service (QoS) guarantees to end customers who are prepared to pay a premium. We refer to this cohort as Differentiated QoS (D-QoS) standards and the aim of this article is to understand the motivations that drove the growth of D-QoS standards, why some have struggled commercially, and draw lessons for future attempts at D-QoS, especially 5G Network Slicing. In this paper, we acknowledge that many D-QoS standards are technically dissimilar. Accordingly, our focus is on the *'goals similarity'* of the D-QoS cohort and not their *'technical similarity'*.

D-QoS standards have a clear purpose in the networking community. S. Keshav notes that "the Holy Grail of computer networking is to design a network that has the flexibility and low cost of the internet, yet offers the end-to-end quality-of-services guarantees of the telephone network" [1]. In practice, however, offering *different* QoS levels to *different* services has been a contentious topic, and has often been associated with concerns about network neutrality. This is because D-QoS is closely associated with the possibility of ISPs selling different Classes of Service (CoS) and the possibility that those who do not pay a premium will receive a poorer experience in contravention of Net Neutrality expectations (where they exist). Accordingly, D-QoS often pitches ISPs against customers, CPs and regulators.

The history of D-QoS standards since the 1970s includes efforts at providing a premium service by treating either some packet flows or the individual packets differently (see Figure 6). The first group includes the use of leased lines; virtual circuits for Asynchronous Transfer Mode (ATM) and Frame Relay; Multi Protocol Label Switching (MPLS), and the combined Integrated Services (IntServ) and ReSerVation Protocol (RSVP). This group mostly implements a connection-oriented mechanism, maintaining a persistent connection for the duration of the connection to mimic reserving a telephone circuit. The second group builds on the packet header concept introduced for IP Type of Service (ToS). It includes Differentiated Services (DiffServ), 4G QoS Class Identifier (QCI), and 5G QoS Indicator (5QI). This group mostly implements a connection-less mechanism, with no mandatory reserved circuits.

D-QoS standards vary widely in their technical design and implementation. This variation has made it almost impossible in the technical literature to compare them. For example, there is very little technical similarity between fixed-line ATM and 4G QCI. Likewise, the end-to-end isolation concept that will be provided by 5G Network Slicing is difficult to compare to most other D-QoS standards. However, from a commercial perspective, all D-QoS standards are eerily similar in seeking to provide a guaranteed connection that customers could be prepared to pay for. In other words, despite their wide technical differences, Leased Lines and 5QI are analysed together as D-QoS standards because they both seek to achieve the similar commercial goal. The purpose of this article is to blend this technical differences, together with the commercial reality, into a framework that can provide insights to both technical and commercial stakeholders.

† Work done while the author was at King's College London, UK.

The concept of Network Slicing in 5G networks is a new development that leverages the inherent softwarisation and virtualisation of 5G networks. Its core idea is to create logical networks over the same physical infrastructure and to then customize and isolate these logical networks according to the requirements of each slice (e.g. different use cases, tenant, virtual operator, etc) [2]. Consequently, from a business perspective, Network Slicing is a means to provide different experiences to different customers, making it similar to classic D-QoS mechanisms. In order words, Network Slicing is considered as a D-QoS in this paper because it has similar goals as other D-QoS and not because it is technically similar to them. This is a key contribution of this article and to the best of our knowledge, this is the first article to do so, and also to link Network Slicing to other historical attempts to offer D-QoS. We highlight that the history of D-QoS suggests that the commercial success of 5G Network Slicing is not as assured as is often assumed [3].

The rest of the article is organised as follows. In Section II, we clarify the scope of this article to keep the analysis compact. Section III provides a summary of selected D-QoS standards, their relative market success, their main advantages and disadvantages. Section IV investigates the technical, regulatory and commercial opportunities and challenges that have impacted these D-QoS standards. In section V, we situate 5G Network Slicing into this category and explore the parallels that can be drawn from previous attempts to offer D-QoS. Section VI provides some concluding remarks.

## II. SCOPE OF PAPER

D-QoS standards - including protocols, technologies and implementations - promise a guaranteed QoS in several layers of the networking stack on both fixed and wireless data networks. In practice, D-QoS mechanisms can be Category 1 (used for operational reasons or within private networks) and Category 2 (designed to be productised and commercialised as a premium service in a pubic network). For example, in 4G (LTE) networks, the ARP (Allocation and Retention Policy) mechanism is an example of Category 1 while QCI is an example of Category 2. In WiFi networks, there are numerous examples of QoS standards, as described in [4] that are examples of Category 1. However, Wireless Multimedia Extension (WME) is an example of a Category 2 standard that have been productised by several vendors (e.g. Aruba), companies (e.g. iPass) and venues (e.g. airports, hotels). This article focuses only on Category 2 standards.

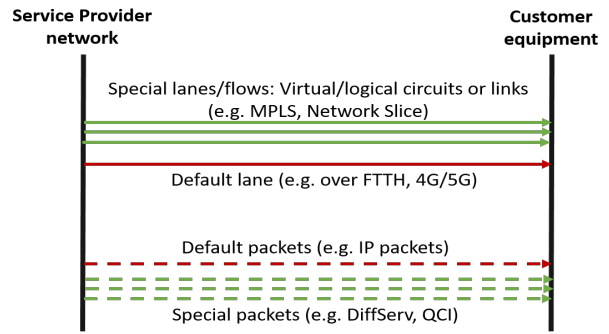

Fig. 1: Differentiated QoS (D-QoS) can be implemented as either special lanes (e.g. via flow/path/route reservation and prioritisation) or special packets (e.g. packet marking)

Using the OSI network model as a guide, and starting with the classification provided by X Xiao in [5], we categorise different D-QoS standards into the layers where they are applicable. In the physical layer, a leased line provides a dedicated, physical connection between two end points. In the datalink layer (Layer 2), examples include QoS provided by X.25, ATM, Frame Relay, Carrier Ethernet and 5G Network Slicing. For the network layer (Layer 3), examples include QoS provided for IP Type of Service (IP TOS), IntServ/RSVP, DiffServ, Software Defined Wide Area Networks (SD-WAN), QCI and 5QI. QoS for MPLS, especially with the traffic engineering components (MPLS TE) is often regarded as an intermediate Layer 2.5 solution. All these D-QoS standards are in Category 2 and in scope for this article.

On the transport layer (Layer 4), while several improvements over TCP and UDP - e.g. Stream Control Transmission Protocol (SCTP), QUIC, Recursive InterNetwork Architecture (RINO) - have been implemented, they are out of scope of this article as they have not been explicitly commercialised as a premium D-QoS service over the 'best effort' connectivity. Similarly, this article will not consider standards across any OSI layer which have not been explicitly offered to customers as a premium service or which are only applicable on networks that are under a single or closed group administrative control. This includes Deterministic Networks (DetNet), Time Sensitive Networking (TSN), X-Ethernet, and the use of MPLS for Mobile Backhaul (as defined in BBF TR-221) etc.

## III. HISTORY OF DIFFERENTIATED QOS STANDARDS

The history of D-QoS standards on data networks can be said to have begun in the 1970s when leased lines were used to connect mainframe computers with terminals and remote sites, offering better reliable connectivity than was obtainable on the telephone network. By 1976, the X.25 option was introduced to provide a global packet switched network. As the popularity of data networks grew in the 1980s, the need for a more scalable option led to the introduction of ATM and Frame Relay networks. Modern variants of these are MPLS and Carrier Ethernet. All of these are connection-oriented, designed to provide special lanes to some packet flows. Commercially, they were offered to the enterprise market, serving a limited number of customers, but contributing sizeable revenues to ISPs. The introduction of IntServ and RSVP attempted to provide a reserved circuit for the mass market access to the internet. On a classic OSI model, these connection-oriented approaches would come under Layer 2 (e.g. ATM, Frame Relay), Layer 3 (e.g. IntServ) or in-between Layer 2/3 (e.g. MPLS as a Layer 2.5 solution).

An alternative approach to offering D-QoS standards on data networks took off with the standardisation of IP ToS, as part of TCP/IP, in 1981. Rather than seek to create a virtual connection, IP ToS used different labels on the headers of data packets to determine the priority to be assigned to each packet. Although it was hardly implemented, its proposal informed the development of DiffServ in 1998. In turn, DiffServ informed the introduction of QCI on 4G cellular networks in 2008. Beginning from 2014, SD-WAN, a new network architecture that brings flexibility of deployment is gaining grounds as a replacement for MPLS in enterprise networks. On a classic OSI model, these connectionless approaches would typically come under Layer 3 - the Network layer.

5G intends to support several mission critical applications (e.g. remote surgery; connected cars; massive numbers of IoT devices, some of which may be monitoring critical infrastructure, etc). Therefore, a flexible and expressive means of ensuring D-QoS is indispensable in delivering these promises. Network Slicing has been suggested as the key mechanism for delivering on this promise, combining isolation and prioritisation of selected traffic flows, a classic D-QoS goal. Table I provides a chronological summary of notable D-QoS standards from the 1970s till date.

## IV. FACTORS THAT DETERMINE SUCCESS FOR DIFFERENTIATED QOS STANDARDS

The basic goal for all D-QoS standards is to provide special treatment to some traffic in a public network. This special treatment can be either superior or inferior to the typical treatment for other traffic. Yet, despite this simplistic goal, there have been very few sustainable success stories (e.g. MPLS) and many more failed attempts at offering D-QoS service classes. Some of the reasons for this have been documented in academic texts, industry articles and even political commentaries. [5] explored in detail the technical, commercial and regulatory challenges of offering D-QoS.

In [6], Benjamin Teitelbaum and Stanislav Shalunov argued that premium IP service has not being commercially deployed and probably never will. In his testimony to the US Senate Committee on Commerce, Science and Transportation hearing on Net Neutrality on 7 February 2006, Gary Bachula, the Vice President of Internet2 said: "For a number of years, we seriously explored various "quality of service" schemes, including having our engineers convene a Quality of Service Working Group. As it developed, though, all of our research and practical experience supported the conclusion that it was far more cost effective to simply provide more bandwidth. With enough bandwidth in the network, there is no congestion and video bits do not need preferential treatment."

While there is merit in these views, we argue that their observation applies mostly to D-QoS that seeks to charge a premium to a mass market audience. This is akin to the 'soft assurance' described in [5] where an ISP sells QoS as a separate entity on top of its regular connectivity service. A broader evaluation of D-QoS approaches shows that services such as leased lines, ATM, carrier ethernet and MPLS have been commercially successful within the limited enterprise space. In the mass market segment, VoLTE is on track to be a commercial success thanks to the regulation of voice services in cellular networks. These approaches are akin to the 'hard assurance' of [5] where the ISP provides predictable QoS *all the time* as part of an already-paid-for service. They also showcase how QoS can be sold as a 'feature' of a product and not as the product itself - in line with the recommendation of [5] to price QoS into the services whilst avoiding to sell QoS explicitly.

Given all the effort that has gone into developing D-QoS standards, a recurring question is why, given its simplistic goal, has it proven difficult to commercialise them as evident in Table I? Based on the different challenges identified in the literature, we identify and cluster 12 key questions that determine the relative success or failure of D-QoS solutions.

### A. Are the technical standards ready?

This is a fairly straightforward requirement and is necessary to avoid fragmentation in effort. It explains why standards bodies are actively involved in creating technical standards for D-QoS services. Examples of the squabbles with standards is the disagreements and redrafting of the different DiffServ standards between 1999 and 2002 that led to RFC2598, RFC3248 and RFC3246 [6]. Ultimately, the D-QoS standards under review in this article have being standardised for implementation.

### B. How simple is the implementation and operations?

While standardisation is the first step to commercialisation, the actual implementation and the day-to-day operations of the service can make or mar its commercial success. Operational considerations include suitability of equipment (e.g. routers), error correction mechanisms, policing of ingress and egress, accounting of different traffic types, etc. An example is X.25 which, as one of the first D-QoS services, was eventually replaced because it was slow and inefficient due to its resource allocation mechanism and error correction procedures. For WME, the lack of consistent tagging of packets has led vendors (e.g. Aruba) to offer WiFi equipment that uses deep packet inspection (DPI) to identify packets for prioritisation.

*C. Does it align with the prevailing ethos of networking?*

As the internet and TCP/IP have become the de facto global standards for data networking, every D-QoS service is judged from the philosophical ethos of the networking community. [7] notes that the fundamental goal of the DARPA project was to find an effective way to integrate existing networks, in a best effort way, such that there was survivability in the face of failure. [6] warns that if premium services were to become common, the best effort traffic will become degraded, radically overhauling the end-to-end design ethos for the internet. BEREC's guidelines for Net Neutrality in the European Union contain the 'necessity' and 'capacity' requirements, forcing ISPs to prove that their D-QoS standards is both necessary and will not degrade the quality of the best effort services.

*D. Are Service Providers free to select users?*

This is at the crux of most of the debate over Net Neutrality. As private companies, ISPs seek the freedom to vary their service offering as they deem fit. But access to the internet is increasingly seen as a 'Public Good' which should be assured for all. We make two clear distinctions about this freedom to select users. Firstly, Table 1 shows that services which are offered to a limited user base (e.g. ATM or MPLS for enterprise customers) are generally permissible and successful compared to services which are aimed at the mass market (e.g. DiffServ)(see fig 7a). Secondly, in cases where regulators believe that society is best served by assuring preferential treatment for a service (e.g. voice), a mass market D-QoS service (e.g. VoLTE on QCI Level 1) has a high probability of commercial success (see fig 7b).

*E. Can customers verify the service?*

It is generally difficult for customers to test and verify the claims of the D-QoS standards. Some authors have used the analogy of jiggling the locks on a door to show that it works correctly [6]. A contrasting analogy is to compare it to insurance services which customers buy - without testing and verifying - because of price, peace of mind, convenience or it was legally mandatory. Enterprise users regularly buy D-QoS services such as ATM and MPLS for similar reasons showing that service verification is not an insurmountable obstacle.

*F. Are Service Providers willing to offer service assurance?*

While there are clear Service Level Agreements (SLAs) for many of the D-QoS services targeted at the enterprise user, ISPs are generally unwilling to provide similar SLAs to the mass market user. [5] observes that offering SLA to the mass market is highly improbable because no one can guarantee anything under a network catastrophe. Without a clear promise of the benefit of using the service, many D-QoS proposals have failed to be implemented.

*G. Is the service simple to sell?*

Selling D-QoS services as a premium service to a mass market audience is an example of 'double selling' where ISPs have to first sell the basic network connectivity service and then sell the premium option on top. In addition, from an ISP's perspective, actively promoting D-QoS options is a tacit admittance of poor service levels with its attendant consequences in a competitive market where ISPs compete on 'network quality'. A convenient scenario exists where an ISP can externalise the poor service level to someone else, or to a faceless entity such as the weather or the Internet. This is often seen in the marketing materials for MPLS, Carrier Ethernet or Leased Lines where ISPs extol the merits of the D-QoS option above the basic internet service that they also provide.

*H. Does the service depend on other Service Providers?*

For an internet that was designed to interconnect diverse networks that are managed independently, it is bound to be difficult to assure end-to-end premium QoS when the service relies on another ISP. That is, an ISP should not make QoS promises to its customers, and then expect a competing ISP to help it in fulfilling that promise without any recompense. [5] contains a copy of a typical 'Peering' contract showing that there is no binding QoS settlement in them. This is unlike the interconnection regime for voice traffic which is underpinned by clear SLAs and a termination fee. This brings up the inter vs intra-domain criteria as a major reason why some D-QoS standards are commercially successful. It holds that when end-to-end QoS can be delivered within a single ISP's domain (e.g. ATM, MPLS) or there is a binding interconnection agreement with enforceable QoS SLAs (e.g. QCI for VoLTE), success is more likely (see fig 7b).

*I. Can the service be scaled easily?*

A feature of the internet is that the intelligence resides at the edge of the network while the equipment in the middle does only basic functions of forwarding packets. This has supported the phenomenal growth of the internet. However, implementing D-QoS standards will necessitate the inclusion of intelligence on all the equipment in the core of the network so that end-to-end QoS guarantees can be achieved. This is a huge burden as it means that an ISP will need to commit serious investment to upgrade its network even before it wins its first premium QoS customer. This point supports the observation that D-QoS services are feasible for a limited market such as enterprises (e.g. MPLS) but cannot scale for a mass market (e.g. IntServ/RSVP) [8] (see fig 7a).

*J. How simple is the business model?*

ISPs need simple business models and clarity on who should pay for the premium service in order to sell D-QoS services. Complexity in the business model means it is difficult to explain it to customers, and also brings additional complexity in the OSS/BSS systems of service providers. In addition, [6] observes that businesses that are built around service assurance separate the 'advertised service' from the 'engineered service' without striving for 100% service reliability. This gives them flexibility to vary the service delivery without breaching levels that trigger penalty. Likewise, deciding who will pay has often caused alarm and triggered Net Neutrality push-backs by businesses who assume that an ISP will ask them to pay. However, content providers often pay to achieve higher QoS for their users. This is the basis of the business model of Content Distribution Networks (e.g. AWS, Akamai).

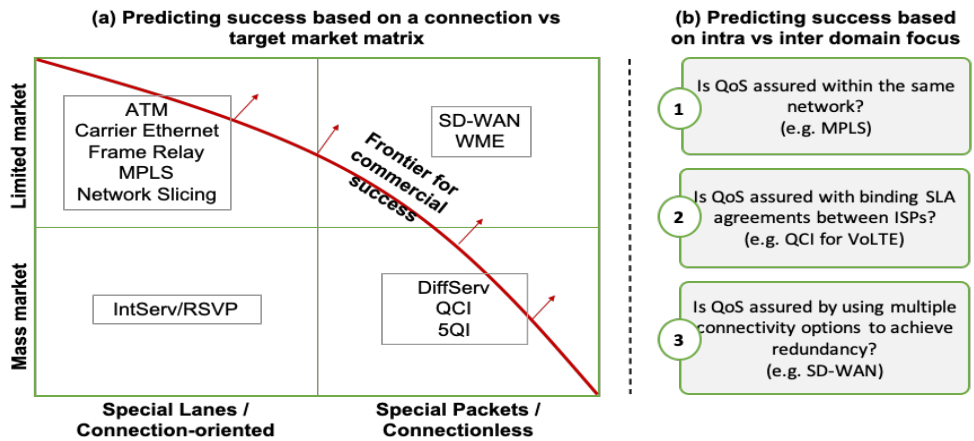

Fig. 2: The history of differential QoS standards highlights how to predict success: (a) Special Lanes / Connection-oriented solutions that are targeted at the mass market are likely to fail (b) Solutions that are delivered across ISP domains, and without a binding SLA, are likely to fail.

*K. Do customers understand the proposition?*

Many of the approaches to sell D-QoS services have been cumbersome and customers have struggled to understand them. This is in line with clear historical evidence that customers prefer simplicity when buying communications services. This is evident from the post 1830s Postal Service and post 1970s telephone service [9], post 2010 fixed broadband service and post 2015 mobile broadband service. The observation is that over a long period of time over a number of industries, pricing for a service is based on either the Service Provider's desire to maximise utilisation and profits or a user's desire for value and simplicity. Inexorably, the trend is towards simplification. A view we have often heard - from speaking to technical and commercial personnel in multiple telecoms operators - on the low-use of the different QCI levels, other than Level 1 for VoLTE, is that customers would not understand the complexities of the differences and so there is no appetite from customers for them. VoLTE benefits from the clarity that across all countries, it is clear to both ISPs and users that voice traffic will be prioritised. [10] suggests a taxonomy to harmonise service prioritisation criteria globally.

*L. Is the service superior to its substitutes?*

For most D-QoS services, the substitute to compete against is a progressively improving best effort connection. For the mass market, customer preference have decidedly been in favour of the best effort offering. Even in enterprise markets where there is convincing case for a premium service, the trend is towards best effort. For example, there is growing evidence that SD-WAN, which can utilise a best-effort internet connection to high QoS levels, is increasingly being preferred to MPLS and leased lines by enterprise customers [11]

V. LESSONS FOR 5G NETWORK SLICING

As the 5G era begins, Network Slicing is being presented as the mechanism for addressing different enterprise requirements, and unlocking commercial opportunities for service providers [2]. Network Slicing promises to deliver a D-QoS in 5G networks by implementing several logical/virtual networks on top of a single 5G network infrastructure [3]. This is an example of a connection-oriented D-QoS, offering a different 'lane' to selected traffic. The other 5G QoS mechanism is 5QI, a connection-less mechanism that builds on 4G QCI to enable different traffic types to be associated with different priority tags. Like QCI, 5QI assigns priority to each individual bearer traffic to ensure that the network can deliver prioritisation for different traffic types.

In this article, we argue that the commercialisation of Network Slicing will ultimately follow the same approach as other D-QoS services, and face the same challenges. While we acknowledge that logical isolation was not a core idea of ATM/MPLS etc, we make a distinction between the *guarantees* provided for QoS, and the *mechanisms* used to provide those guarantees. Our thesis rests on the former rather than the latter, illustrating a similarity between Network Slicing and many other D-QoS services, which has not being explored in the literature. In this section, we assess

Network Slicing using the same parameters and summarise the analysis in Figure 8.

## A. Standards for Network Slicing

3GPP Release 15 has delivered the first tranche of the standards for Network Slicing and there have been several proof-of-concepts in industry and academia demonstrating that the standards are implementable.

According to [12], 3GPP defines slices as a "logical network that provides specific network capabilities and network characteristics." Network slices provide a functional construct to achieve two orthogonal objectives: (i) isolation between traffic which may interfere with each other; and (ii) a network design which best supports the mix of applications that are running within the slice. Generally, [3] describes three solution groups that are discussed with varying levels of common functionality in 3GPP standards: Group A is characterised by a common Radio Access Network (RAN) and completely dedicated Core Network (CN) slices, i.e., independent subscription, session, and mobility management for each network slice handling the UE. Group B also assumes a common RAN, where identity, subscription, and mobility management are common across all network slices, while other functions such as session management reside in individual network slices. Group C assumes a completely shared RAN and a common CN control plane, while CN user planes belong to dedicated slices.

Because each slice can be tailored to support a particular application or mix of applications, network slices can be used to provide strong QoS guarantees (based on isolation from other network slices). Taken a step further, [13] introduces the concept of Network Slicing as a Service, demonstrating how it can be used to enable enterprises with their own software-defined cellular networks.

## B. Implementation for Network Slicing

Building on the 3GPP standards, ongoing activities at the GSMA and other industry fora are focused on providing implementation guidelines for Network Slicing, especially with a Generic Slice Template (GST) - created by the GSMA in 2018 to minimise the risk of a fragmented implementation across the industry. These will help, but are unlikely to address all implementation challenges or completely eliminate the risk of fragmented implementation globally.

## C. The philosophy of Network Slicing

Network Slicing effectively treats different service types differently, raising questions on if this will be acceptable under net neutrality principles [14]. However, given that it is generally accepted to offer differentiated services to enterprise customers using leased lines and MPLS, we posit that there is unlikely to be sustained opposition to Network Slicing based on net neutrality rules.

## D. Customer selection for Network Slicing

As a service targeted to enterprises, ISPs should be able to apply the same commercial framework that applies to existing and historical D-QoS standards (e.g. MPLS, Leased Lines and ATM) to Network Slicing.

## E. Service assurance for Network Slicing

ISPs will be under pressure to offer binding SLAs to enterprises to convince them to adopt slicing. This should be expected as enterprises have stringent business outcomes to achieve.

## F. Customer verification of Network Slicing

Enterprise customers generally buy D-QoS standard solutions, without actively verifying the service, as an insurance policy against adverse business outcomes. The same should apply to Network Slicing.

## G. Simplicity of selling Network Slicing

This is currently unclear and would be tested in the market as Standalone 5G networks become available. However, many ISPs already have expertise selling MPLS/leased lines to enterprises and will have to transfer those selling skills to Network Slicing.

## H. End-to-end interconnected Network Slicing

As demonstrated by Kings' College London, working with BT and Ericsson in 2018, interconnecting slices from different ISPs to create a global end-to-end slice is feasible [1]. However, achieving seamless interoperability and interconnection would be a challenge and it would always be easier to deliver slices within a single ISP's network domain.

## I. Customer understanding of Network Slicing

Effort is being made via academic and industry papers, seminars and conferences to sensitise customers about the possibilities with Network Slicing (e.g. [15]). This will become clearer as standalone 5G networks are deployed.

## J. Scalability of Network Slicing

There is ongoing debate as to how many slices will be created per ISP. In a July 2018 blog post titled "How many network slices are needed?", Ericsson argued that the more autonomous slices can be - and the ability to automate their management - will determine how many slices an ISP can cost-effectively deploy.

## K. Simplicity of Network Slicing

This is currently unclear and will become more evident as Network Slicing is rolled out in standalone 5G networks

---

[1]https://www.ericsson.com/en/press-releases/2017/2/ericsson-and-bt-put-5g-to-work-in-real-world-applications

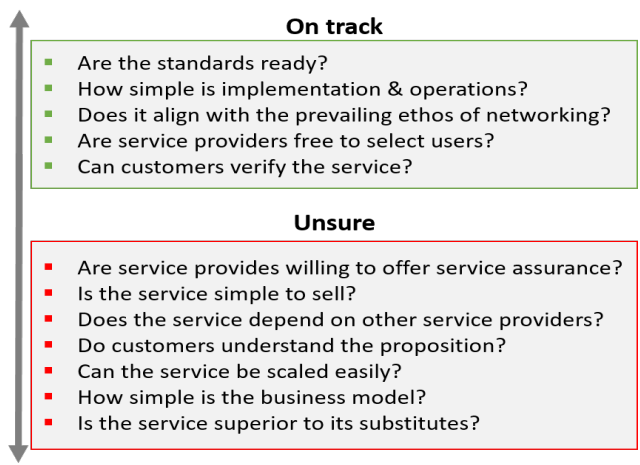

Fig. 3: Evaluating Network Slicing based on lessons from previous differentiated QoS services

*L. Alternatives to Network Slicing*

There are three competing alternatives to Network Slicing. First, the improving connectivity from 5G could encourage some potential enterprise customers to decline using dedicated slices. Secondly, some customers may prefer using SD-WAN to slices. Third, for customers who need isolation or hands-on control of their network infrastructure, a personalised, private 5G network is an alternative. Quoting industry sources, a ComputerWeekly article in April 2019 predicted that such private 5G networks will be a 'big trend'.

*M. Summary: Recommendation for Network Slicing*

From our review of how all the factors may impact Network Slicing, it is clear that the success of Network Slicing will depend on much more than technical factors. However, much of the academic and industry literature on the challenges of Network Slicing has so far focused on the technical challenges associated with standardisation and implementation [15]. Yet, those are only 2 (Points 1 and 2) out of the 12-point checklist that will determine the success or failure of Network Slicing as shown in the previous section.

Given the comparative analysis on Network Slicing in this section with the generic observations about other D-QoS (Section IV), plus the key insights shown in Figure 7, we conclude that the possibilities of success can be high if attention is given to clarifying the business model and suggest that as an enterprise-focused proposition, Network Slicing has a strong chance of success if it is provided within an ISPs domain only, and with binding SLAs.

## VI. CONCLUSION

The history of D-QoS standards on data networks shows that there have been several success stories, especially when the target market is limited and the service assurance is provided inside an intra-domain network. Targeting the limited enterprise market has been a fairly good predictor of success whereas aiming to reach out to the mass market has largely failed to deliver commercial success. However, even in the enterprise space, the improving quality of the best effort internet is leading increasingly to a rethink by IT managers on the merits of paying for expensive D-QoS services. This has implications for 5G Network Slicing as it seeks to debut in a market where the appetite for D-QoS services that are priced at a premium is diminishing. While work is ongoing to finalise the technical standards and operational details of Network Slicing, the lessons from the history of existing and previous standards suggest that if it is to be equally successful, there is a need for better clarity on the business model and also the level of service assurance that will be provided to users.

## VII. BIOGRAPHY

*A. Emeka Obiodu*

Emeka Obiodu (chukwuemeka.obiodu@kcl.ac.uk) is a PhD student in the Department of Engineering at King's College London where his research focuses on differentiated services in the 5G era. His other affiliation is with GSMA, where he is the Strategy Director for Networks & Platforms and the Project Lead for GSMA's 5G Taskforce. Previously, he spent 10 years as a telecoms analyst/consultant, covering the introduction and evolution of 3G & 4G markets globally. He has a BEng from Federal University of Technology Owerri (FUTO), an MSc in Telecoms from Queen Mary University of London, and an MBA from University of Warwick Business School.

*B. Nishanth Sastry*

Nishanth Sastry(n.sastry@surrey.ac.uk) is a Professor of Computer Science at University of Surrey, UK. Previously, he spent nine years at King's College London (Departments of Engineering/Informatics), and over six years in the Industry (Cisco and IBM). He has been a visiting researcher at the Alan Turing Institute and Massachusetts Institute of Technology. He holds a Bachelor's degree (with distinction) from R.V. College of Engineering, Bangalore University, a Master's degree from University of Texas, Austin, and a PhD from University of Cambridge, all in Computer Science. He has been granted nine patents in the USA for work done at IBM.

TABLE I: Selected history of differentiated QoS Standards: 1970s till date

| Name | Description | OSI layer & Connection type | Standards owner & year | Current market status & value | Main benefit | Main constraint |
|---|---|---|---|---|---|---|
| Leased Lines | Provides a virtual, constant connection between two end points | Layer 2; Special lanes; limited market | ITU; 1974 | Moderate usage; moderate revenues (Ofcom UK) | Provides guaranteed connectivity with clear SLAs | Too expensive for most users |
| X.25 | A global packet switched network with rigorous error correction mechanisms and based on the PSTN paradigm | Layer 2; Special lanes; limited market | ITU; 1976 | Little evidence of current usage | Provides QoS and error free delivery over links of high error-rate | Slow and inefficient due to resource allocation mechanism and error correction |
| IP TOS | The first conceptual provision for QoS on IP networks. Uses an 8-bit field to indicate delay, throughput and reliability | Layer 3; Special packets; mass market | IETF; 1981 | Little evidence of current usage | Conceptualised QoS for IP networks | No real need in early IP networks because of unreliable connections |
| ATM | An approach to recreate PSTN guarantees on a data network by providing 'virtual' circuits. Supports resource reservation and admission control with decent performance guarantees | Layer 2; Special lanes; limited market | ITU; 1988 | Decreasingly used; negligible revenues | Provides QoS guarantees in addition to allowing resource sharing for both high throughput traffic and real time traffic | High, and increasingly irrelevant complexity leading to high cost |
| Frame Relay | Transmits data in variable-length 'frames' while maintaining a virtual circuit between end nodes | Layer 2; Special lanes; limited market | ITU; 1990 | Decreasingly used; negligible revenues | Provided QoS guarantees without the cost of dedicated/leased lines and without the complexity of X.25 | Slower than ATM with a similar cost burden |
| IntServ / RSVP | The first real attempt to implement QoS on IP networks. Reserves the private circuits akin to the telephone using RSVP | Layer 3; Special lanes; mass market | IETF; 1994 | Little evidence of current usage | Proof of concept for QoS | Too complex and not scalable |
| Diffserv | Proposed to overcome the scalability constraint of IntServ and builds on IP TOS | Layer 3; Special packets; mass market | IETF; 1998 | Little evidence of current usage | Good for real time traffic (e.g. VoIP) | Needs every node on the path to support |
| MPLS | Provides a mechanism for setting up a connection-oriented path over different network types | Layer 2.5; Special lanes; limited market | IETF; 2001 | Widely used; $46.3 billion by 2020 (Grandview Research) | Provides QoS guarantees over an IP network by setting up paths prior to sending packets through the network | Costly for many enterprises and the improvement in internet quality questions the relevance of MPLS' QoS guarantees |
| Carrier Ethernet | An extension of the LAN into a WAN. It provides a mechanism to circumvent bandwidth bottlenecks that can occur when a large number of small networks are connected to a single larger network | Layer 2; Special lanes; limited market | MEF; 2001 | Widely used; $22.5 billion by 2020 (Ovum) | Integrates neatly into the LAN as an all-Ethernet network | No integral support for differential QoS. Limited scale compared to MPLS. As such, mostly used for connecting data centres |
| WME | Provides priority to selected traffic (e.g. voice) over IEEE802.11 networks | Layer 2; Special packets; limited market | IEEE; 2005 | Integrated into most enterprise WiFi | Delivers QoS to VoIP/video services on congested WiFi networks | Initial tagging of packets is often missing, necessitating vendor workarounds with deep packet inspection |
| QCI | Based on the DiffServ model and is the first QoS model for a fully packet-based cellular network | Layer 3; Special packets; mass market | 3GPP; 2008 | VoLTE use worth $34.8 billion by 2022 (AM Research) | Adapts the DiffServ QoS paradigm to cellular services | Needs every ISP on the path to support. Hence commercially unattractive |
| SD-WAN | Moves the control of WAN networks into the cloud whilst providing high QoS at a low cost over the public internet | Layer 3; Special packets;, limited market | 2014 | Growing usage; $1.3 billion by 2021 (Gartner) | Provides the QoS guarantees of MPLS at a fraction of the cost | Unlikely to match the QoS guarantees of MPLS for some business critical services |
| 5QI | Builds on QCI and will be the de-facto QoS standard to support ultra reliable and low latency 5G services | Layer 3; Special packets; mass market | 3GPP; 2017 | Not yet launched | Designed to support ultra reliable and low latency services | Commercially unproven |
| Network Slicing | Mechanism to create multiple logical/virtual networks on top of a single physical network | Layer 2; Special lanes; limited market | 3GPP; 2017 | Not yet launched | Designed to provide connection-oriented QoS guarantees over cellular data networks | Commercially unproven |